\documentclass[smallextended]{svjour3}       
\smartqed  
\usepackage{graphicx}
\usepackage{amsmath}
\journalname{Journal of Low Temperature Physics}
\begin{document}

\title{Molecular Tagging Velocimetry in Superfluid Helium-4: Progress, Issues, and Future Development}

\author{Wei Guo}


\institute{W. Guo \at
              Florida State University \\
              Mechanical Engineering Department\\
              National High Magnetic Field laboratory\\
              1800 East Paul Dirac Drive\\
              Tallahhassee, Florida 32310, U.S.A.
              \email{wguo@magnet.fsu.edu}
}
\date{Received: date / Accepted: date}
\maketitle

\begin{abstract}
Helium-4 in the superfluid phase (He II) is a two-fluid system that exhibits fascinating quantum hydrodynamics with important scientific and engineering applications. However, the lack of high-precision flow measurement tools in He II has impeded the progress in understanding and utilizing its hydrodynamics. In recent years, there have been extensive efforts in developing quantitative flow visualization techniques applicable to He II. In particular, a powerful molecular tagging velocimetry (MTV) technique, based on tracking thin lines of He$^*_2$ excimer molecules created via femtosecond laser-field ionization in helium, has been developed in our lab. This technique allows unambiguous measurement of the normal-fluid velocity field in the two-fluid system. Nevertheless, there are two limitations of this technique: 1) only the velocity component perpendicular to the tracer line can be measured; and 2) there is an inherent error in determining the perpendicular velocity. In this paper, we discuss how these issues can be resolved by advancing the MTV technique. We also discuss two novel schemes for tagging and producing He$^*_2$ tracers. The first method allows the creation of a tagged He$^*_2$ tracer line without the use of an expensive femtosecond laser. The second method enables full-space velocity field measurement through tracking small clouds of He$^*_2$ molecules created via neutron-$^3$He absorption reactions in He II.
\end{abstract}

\section{Introduction}\label{intro}
Liquid helium-4 ($^4$He) transits to the superfluid phase (known as He II) below about 2.17 K \cite{Tilley-book}. In He II, two miscible fluid components co-exist: an inviscid superfluid (i.e., the condensate) and a viscous normal fluid (i.e., the thermal excitations). The hydrodynamics of He II is strongly affected by quantum effects. For instance, the rotational motion of the superfluid component can occur only with the formation of topological defects in the form of quantized vortex lines \cite{Donnelly-book-1991}. These vortex lines all have identical cores with a radius of about 1 {\AA}, and they each carry a single quantum of circulation $\kappa=h/m_4$, where $h$ is Planck¡¯s constant and $m_4$ is the mass of a $^4$He atom. Turbulence in the superfluid therefore takes the form of an irregular tangle of vortex lines (quantum turbulence) \cite{Vinen-JLTP-2002}. The normal fluid is expected to behave more like a classical fluid. But a force of mutual friction between the two fluids \cite{Vinen-PRS-1957}, arising from the scattering of thermal excitations by the vortex lines, can affect the flows in both fluids.

The fraction ratio of the two fluid components in He II strongly depends on temperature. Above 1 K where both fluids are present, this two-fluid system exhibits fascinating hydrodynamic properties that have important scientific and engineering applications \cite{Sciver-book-2012}. For instance, He II supports the most efficient heat-transfer mechanism (i.e., thermal counterflow) and therefore has been widely utilized for cooling scientific and industrial equipment such as superconducting magnets, particle colliders, superconducting accelerator cavities, and satellites \cite{Blanco-AIP-2004,Taber-AIP-2002,Bonitooliva-Cryo-1994}. It has also been suggested that He II can be used to generate flows with extremely high Reynolds numbers for model testing of large-scale classical turbulence that can hardly be achieved with conventional test fluids such as water and air \cite{Donnelly-HiRe-1991,Skrbek-JP-1999,Sreenivasan-AAM-2001}. However, despite decades of research, the full potential of He II has not yet been realized, largely due to the lack of high-fidelity quantitative flow measurement tools.

Typical single-point diagnostic tools used for classical fluids, such as pitot pressure tubes and hot-wire anemometers, either have limited spatial resolution or rely on convective heat transfer that does not exist in He II. Furthermore, since the motion of both fluid components can contribute to the sensor response, data analysis can become very complicated when the two fluids have different velocity fields. More straightforward velocity measurements can be made via direct flow visualization \cite{Guo-PNAS-2014}. In the past, researchers used micron-sized solidified particles as tracers and developed particle image velocimetry (PIV) and particle tracking velocimetry (PTV) techniques for He II \cite{Sciver-JLTP-2007,Zhang-NP-2005,Mantia-JFM-2013,Paoletti-JPSJ-2008,Bewley-Nat-2006,Paoletti-PRL-2008}. These micron-sized tracers can easily get trapped on quantized vortices due to their large binding energy to the vortex cores \cite{Guo-PNAS-2014}. The trapped tracers have yielded very interesting images of the vortex lines \cite{Bewley-Nat-2006,Paoletti-PRL-2008,Bewley-PNAS-2008,Fonda-PNAS-2014}. Nevertheless, some issues are known to exist. First, these tracer particles are produced by injecting room temperature gas mixture to liquid helium, which introduces a large heat load and hence disturbs the flow to be studied. Second, the particles produced by this method have a wide range of sizes and irregular shapes and are not neutrally buoyant, which lead to complicated tracer behavior. The strong interaction of the tracers with both the viscous normal fluid and the quantized vortices sometimes makes their motion hardly analyzable in practical turbulent flows \cite{Kivotides-PRB-2008}. Nevertheless, we note that in our recent PTV study of thermal counterflow in He II, a new method for separately analyzing the particles trapped on vortices and those entrained by the normal fluid approves to be very useful \cite{Mastracci-PRF-2018}.

On the other hand, the feasibility of using He$^*_2$ excimer molecules as tracers in He II has been validated through a series of experiments \cite{Guo-PRL-2009,Guo-JLTP-2010,Guo-PRL-2010}. These molecules can be created easily as a consequence of ionization or excitation of ground state helium atoms \cite{Benderskii-JCP-2002}. They have exceptionally long radiative lifetime in the electron spin triplet state (about 13 s \cite{McKinsey-PRA-1999}) and form tiny bubbles in liquid helium (about 6 {\AA} in radius \cite{Benderskii-JCP-2002}). Due to their small size and hence small binding energy on vortex cores \cite{Mateo-JCP-2015}, trapping of the He$^*_2$ tracers by quantized vortices can occur only below about 0.6 K in the absence of the normal fluid \cite{Zmeev-PRL-2013}. At above 1 K where most of the He II based applications take place, He$^*_2$ tracers are solely entrained by the normal fluid since the viscous drag force dominates other forces. These He$^*_2$ tracers can be imaged via a cycling-transition laser-induced fluorescence (LIF) technique that was developed by McKinsey's group at Yale where the author worked as a postdoctoral researcher \cite{Guo-PRL-2009,McKinsey-PRL-2005,Rellergert-PRL-2008}. More recently, a powerful molecular tagging velocimetry (MTV) technique based on tracking thin lines of He$^*_2$ tracers, created via femtosecond laser-field ionization in helium, has been developed in our lab \cite{Gao-RSI-2015}. This technique allows for quantitative measurement of the normal-fluid velocity field in the two-fluid system. The application of the MTV technique to thermal counterflow turbulence in He II has yielded remarkably fruitful results \cite{Marakov-PRB-2015,Gao-JETP-2016,Gao-PRB-2016,Gao-JLTP-2017,Gao-PRB-2017,Gao-PRB-2018}.

Nevertheless, there are two obvious limitations of this MTV technique: 1) only the velocity component perpendicular to the tracer line can be measured; and 2) there is an inherent error in determining the perpendicular velocity. In this paper, we will discuss how these issues can be resolved by creating complex tracer-line patterns for advanced MTV measurement. We will also discuss two novel schemes for tagging and producing He$^*_2$ tracers. The first method is an inverse tagging velocimetry scheme without the use of any expensive femtosecond lasers. This method utilizes the vibrational levels of the He$^*_2$ molecules but with much improved tagging efficiency compared to the method reported in ref. \cite{Guo-PRL-2009}. The second method is to create small clouds of He$^*_2$ molecules via neutron-$^3$He absorption reactions in He II. Each small He$^*_2$ cloud can be treated as a single ``tracer'' such that unambiguous PTV or PIV measurements of the normal-fluid velocity field can be performed.

\section{He$^*_2$ molecular tagging velocimetry: issues and solutions} \label{sec:1}
\begin{figure}
\center
  \includegraphics[width=0.8\linewidth]{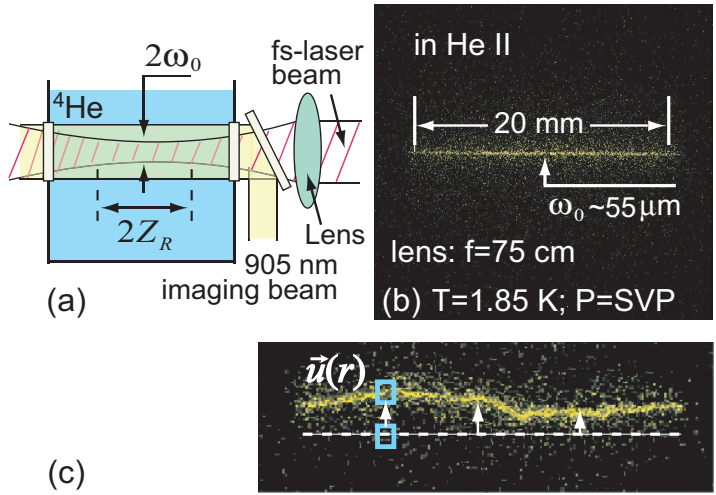}
\caption{(a) Schematic diagram showing the optical setup for creating and imaging He$^*_2$ molecular tracer lines in helium. (b) A typical tracer image in He II upon its creation. (c)Schematic showing how the local velocity can be calculated. The dashed line indicates the tracer line initial location. }
\label{Fig1}       
\end{figure}

To create a line of He$^*_2$ tracers via laser-field ionization, laser intensity as high as 10$^{13}$ W/cm$^2$ is needed \cite{Benderskii-JCP-2002}. This high instantaneous laser intensity can be achieved by focusing a femtosecond laser pulse in helium. In our lab, a regenerative amplifier laser system (wavelength $\lambda$: 780 nm, duration: 35 fs, pulse energy: up to 4 mJ) has been utilized for this purpose. As shown schematically in Fig.~\ref{Fig1} (a), we focus the fs-laser beam using a lens with a focal length $f$ and pass the beam through an optical cryostat that contains He II at regulated pressures and temperatures. Anti-reflection coated windows are used to minimize laser heating. For an ideal Gaussian beam with a beam radius $\omega_0$ at the focal plane, one can define a Rayleigh range $Z_R=\pi{\omega_0}^2/\lambda$, over which the laser intensity drops by 50\% due to beam spreading. The He$^*_2$ tracers are expected to be produced essentially within the Rayleigh range. In our experiment, a fs-laser pulse energy of about 60 $\mu$J is sufficient to create a thin line of He$^*_2$ tracers. We then send in several pulses from a 1-kHz imaging laser at 905 nm to drive the tracers to produce fluorescent light for line imaging. Fig.~\ref{Fig1} (b) shows a typical fluorescence image of the He$^*_2$ tracer line taken right after its creation in He II. The width of the tracer line is about $2\omega_0$ and its length is about $2Z_R$ as expected.

To extract velocity information, we allow an initially straight tracer line to move with the fluid by a drift time $\triangle t$. The deformed tracer line is then divided into small segments so that the center of each segment can be determined by a Gaussian fit of its intensity profile. When the drift time is small, the velocity component $u(r)$ perpendicular to the tracer line can be calculated as the vertical displacement of the line segment at $r$ divided by $\triangle t$. Valuable flow field information, such as the streamwise velocity profile and transverse velocity correlations, can be extracted from $u(r)$. This method has been adopted in various MTV experiments in classical fluids \cite{Miles-FDR-1991,Miles-ARFM-1997}, and it has also allowed us to obtain valuable insights in He II thermal counterflow \cite{Marakov-PRB-2015,Gao-JETP-2016,Gao-PRB-2016,Gao-JLTP-2017,Gao-PRB-2017,Gao-PRB-2018}.

\begin{figure}
\center
  \includegraphics[width=0.99\linewidth]{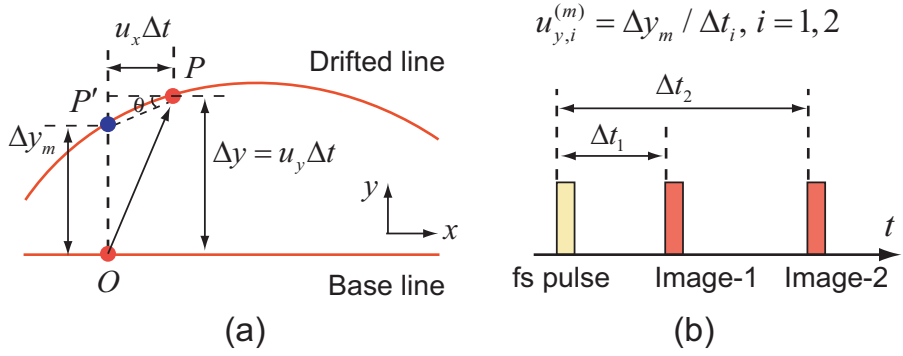}
\caption{(a) Schematic diagram showing the inherent error in determining the velocity component perpendicular to the tracer line. (b) Schematic diagram of the double-exposure method for correcting the inherent error.}
\label{Fig2}       
\end{figure}

Nevertheless, some limitations have been identified. First, by tracking a single tracer line, we can only determine the velocity component perpendicular to the tracer line, from which only the transverse velocity correlation at the line location can be extracted. In turbulence research, it is desirable to have the capability of mapping out the complete velocity field. Second, even for the velocity component perpendicular to the tracer lines, it is known that there is an inherent error involved in the measurement which can become non-negligible when the flow parallel to the tracer line is sufficiently strong \cite{Hammer-MST-2013}. The first issue can be easily understood. To see the second issue more clearly, let us consider the schematic shown in Fig. \ref{Fig2} (a). An initially straight tracer line created parallel to the $x$-axis (i.e., the red solid horizontal line) deforms as it drifts with the fluid by a drift time $\triangle t$. If one is to measure the vertical velocity $u_y$ at point $O$, our existing MTV method will yield a measured velocity $u^{(m)}_y$ that is given by $u^{(m)}_y=\triangle{y_m}/\triangle{t}$, where $\triangle{y_m}$ is the apparent vertical displacement of point $O$. However, when there is a finite horizontal flow, the fluid particle originally located at point $O$ will move to point $P$ instead $P'$. Therefore, the actual vertical velocity at point $O$ should be $u_y=\triangle{y}/\triangle{t}$. The error in the vertical velocity $\triangle{u_y}=u_y-u^{(m)}_y$ is thus given by
\begin{equation}\label{Eq1}
\triangle{u_y}=\frac{\triangle{y}-\triangle{y_m}}{\triangle{t}}=\frac{u_x\triangle{t}\cdot \tan\theta}{\triangle{t}}=u_x\cdot\left(\frac{\partial u_y}{\partial x}\right)\cdot\triangle{t}.
\end{equation}
One sees clearly that the error in the measured vertical velocity is proportional to both the drift time $\triangle{t}$ and the horizontal velocity $u_x$. For a given drift time in an experiment, this error is negligible only if the horizontal flow is small.

To fix this issue, Hammer \emph{et al}. proposed a multi-time-delay approach \cite{Hammer-MST-2013}. Here, we shall discuss a simplified double-exposure version. As illustrated in Fig. \ref{Fig2} (b), following the creation of a tracer line, we may send two imaging pulse trains at drift times of $\triangle{t}_1$ and $\triangle{t}_2$, respectively. The camera will be synchronized with these two imaging pulse trains to capture two corresponding images of the tracer lines. For any given point $O$ on the base line, we can now calculate two vertical velocities using the two images of the deformed line as $u^{(m)}_{y,1}=\triangle{y_{m,1}}/\triangle{t_1}$ and $u^{(m)}_{y,2}==\triangle{y_{m,2}}/\triangle{t_2}$. Note that according to Eq.~\ref{Eq1}, the actual vertical velocity $u_y$ can be evaluated as
\begin{equation}\label{Eq2}
u_y=u^{(m)}_{y,1}+a\cdot\triangle{t_1} \texttt{  and  } u_y=u^{(m)}_{y,2}+a\cdot\triangle{t_2}
\end{equation}
where the parameter $a=u_x\cdot\left(\partial{u_y}/\partial{x}\right)$. From Eq.~\ref{Eq2}, one can then determine the values for both $u_y$ and $a$ at the point $O$. Therefore, the inherent error in measuring $u_y$ can be remedied. Note that this method assumes that the value of $a$ does not change much between the two image acquisitions. This assumption holds when both $\triangle{t_1}$ and $\triangle{t_2}$ are small.

\begin{figure}
\center
  \includegraphics[width=0.99\linewidth]{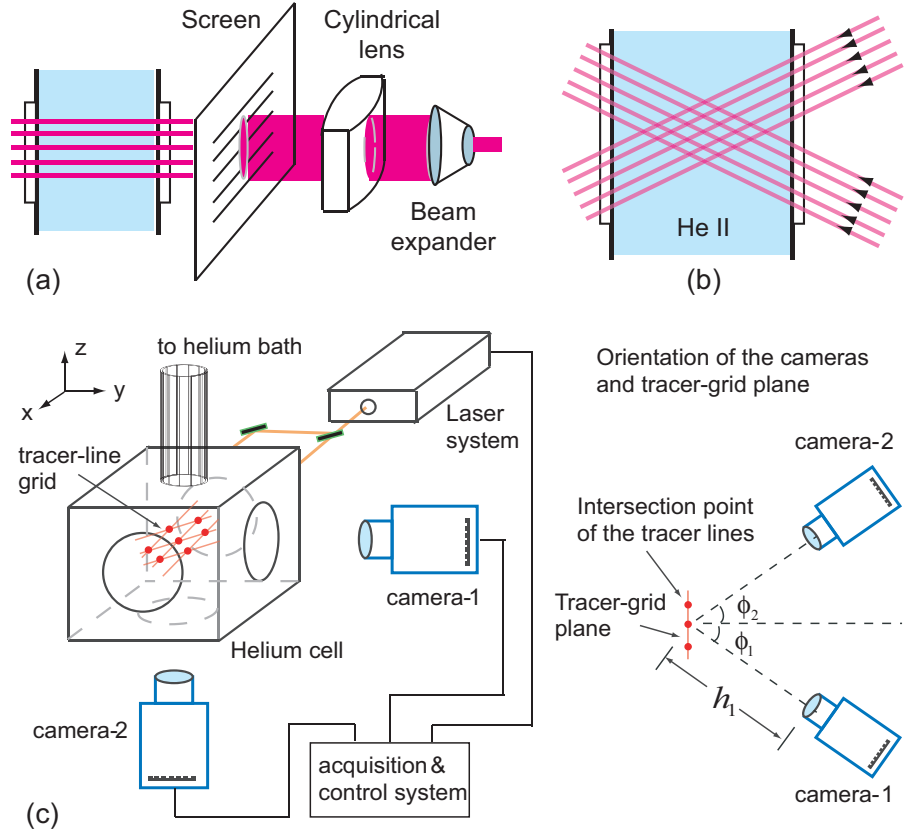}
\caption{Schematic diagrams of the optical systems for producing: (a) a tracer-line array and (b) a tracer-line grid. (c) Schematic diagram showing the experimental setup for stereoscopic MTV measurement in helium.}
\label{Fig3}       
\end{figure}

In principle, based on the definition of $a$ and the obtained $u_y(x)$, one can also calculate the horizontal velocity $u_x$. However, this calculation involves the evaluation of spatial derivative $\partial{u_y}/\partial{x}$ from $u_y(x)$, which is normally highly noisy and undesired. A more feasible route to obtain information about other velocity components is to produce and track complex tracer-line patterns \cite{Chen-AMSP-2015}. Creating multiple tracer lines simultaneously in He II is relatively straightforward using our laser system. The maximum pulse energy of our femtosecond laser (i.e., 4 mJ) is far greater than necessary for creating a single tracer line (i.e., 60 $\mu$J \cite{Gao-RSI-2015}). Therefore, we can divide the fs-laser beam into multiple beams to produce multiple tracer lines that form patterns. Fig. \ref{Fig3} (a) shows the concept how the fs-laser beam can be focused into a thin laser sheet which can then be passed through a screen with many parallel thin open slots for producing an array of tracer lines. Overlapping two such arrays can lead to the formation of a tracer-line grid as shown in Fig. \ref{Fig3} (b). This method has already been adopted by Hu and his colleagues in classical MTV experiments \cite{Chen-AMSP-2015}. 

In order to obtain full three dimensional (3D) velocity information, a stereoscopic imaging system needs to be implemented. Instead of tracking the tracer lines themselves, one may track the motion of the intersection points of a tracer-line grid. These intersection points can be treated as individual ``tracers'' whose velocities can be accurately measured and are not be affected by the typical MTV inherent error. An example stereoscopic MTV setup is shown in Fig. \ref{Fig3} (c). A tracer-line grid can be created using the method as illustrated in Fig. \ref{Fig3} (b). Then, two intensified CCD (ICCD) cameras will be synchronized to take images of the tracer-line grid at the same time from two directions. In a drift time $dt$, the apparent displacements of the intersection points of the grid-lines (i.e., those solid red dots in Fig. \ref{Fig3}) in the imaging planes of the two cameras can be recorded (i.e., ($dx_1$, $dy_1$) and ($dx_2$, $dy_2$)), from which the actual $x$, $y$, $z$ displacements of the intersection points can be computed as:
\begin{equation}\label{Eq3}
\left.\begin{aligned}
dx&=f_x\left(dx_1,dy_1,\phi_1,h_1,m_1;dx_2,dy_2,\phi_2,h_2,m_2\right)\\
dy&=f_y\left(dx_1,dy_1,\phi_1,h_1,m_1;dx_2,dy_2,\phi_2,h_2,m_2\right)\\
dz&=f_z\left(dx_1,dy_1,\phi_1,h_1,m_1;dx_2,dy_2,\phi_2,h_2,m_2\right)
\end{aligned}\right\}
\end{equation}
where $\phi$, $h$, and $m$ are the camera viewing angle, the distance between the image and the object plane, and the image magnification factor. The forms of the functions $f_x$, $f_y$, and $f_z$ are complex but have been discussed in great detail in the literature \cite{Bohl-EF-2001}. The three velocity components can therefore be calculated as $u_x=dx/dt$, $u_y=dy/dt$, and $u_z= dz/dt$. This velocity measurement is free from any ambiguity associated with the MTV inherent error.

\section{Novel schemes for tagging and producing He$^*_2$ tracers}\label{sec:2}
\subsection{Inverse molecular tagging velocimetry}
The MTV technique discussed in the previous section requires a powerful femtosecond laser system for producing He$^*_2$ tracer lines. Such a femtosecond laser system is very expensive and normally not accessible to many research groups who would like to conduct MTV measurements in helium. On the other hand, there are many other ways to produce He$^*_2$ molecules in helium without involving a femtosecond laser. For instance, one may use a radioactive source \cite{Rellergert-PRL-2008} or apply a high voltage on a sharp tungsten needle to ignite a field emission \cite{Guo-PRL-2009,Guo-PRL-2010}. The He$^*_2$ molecules produced by these methods can disperse in the whole fluid. Then, a key question is how to tag or select a group of the dispersed He$^*_2$ molecules at a desired location for imaging so that quantitative velocity information can be obtained by tracking the tagged molecules. Ref. \cite{Guo-PRL-2009} reported a tagging method that utilizes the long-lived vibrational levels of the He$^*_2$ triplet ground state $a^3\Sigma^+_u$. The basic concept is to use a focused nanosecond pump laser pulse at 910 nm to excite a line of He$^*_2$ molecules from their ground state $a(0)$ to the excited electronic state $c(0)$. About 4\% of the $c(0)$ molecules non-radiatively decay in a few nanoseconds to the long-lived vibrational level $a(1)$ and are tagged. An expanded nanosecond probe laser at 925 nm can then be used to only drive the tagged $a(1)$ molecules to the excited $d$ state to induce 640 nm fluorescence via the $d$ to $b$ transition. This method has allowed the first tagging of a He$^*_2$ tracer line in He II thermal counterflow \cite{Guo-PRL-2010}. Nevertheless, due to the low tagging efficiency (i.e., about 4\%), many images must be superimposed at a given pump-probe delay time to achieve a good image quality. Consequently, deformations of individual lines that contain turbulent velocity field information are completed smoothed out.

\begin{figure}
\center
  \includegraphics[width=0.99\linewidth]{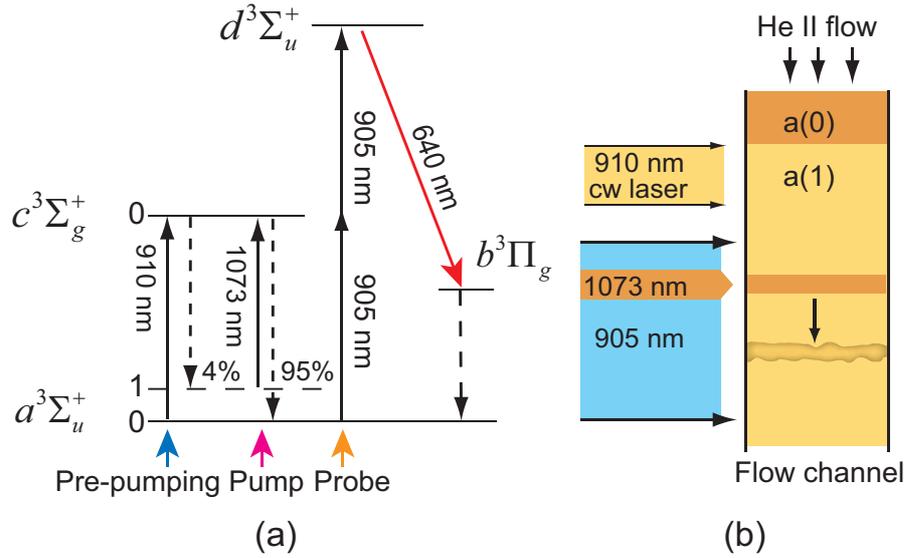}
\caption{(a) Schematic diagram showing the optical transitions of the inverse tagging method. The levels labeled by 0 and 1 for each electronic state are the vibrational levels of the corresponding state. (b) Schematic diagram of the experimental setup that utilizes the inverse tagging method for imaging a single tagged He$^*_2$ tracer line.}
\label{Fig4}       
\end{figure}

Here we would like to discuss a new inverse-tagging method that allows for high quality imaging of individual tagged tracer lines. The scheme of this method is shown in Fig. \ref{Fig4} (a). Instead of tagging the molecules by driving them to the a(1) state, a pre-pumping continuous laser at 910 nm can be used to drive the molecules from the $a(0)$ to the $c(0)$ state. Since the $c$ to $a$ decay time is only a few tens of nanoseconds, molecules that decay back to $a(0)$ can quickly be re-excited to $c(0)$. Each time the molecules promoted to the $c(0)$ state have about 4\% of the chance to decay to the $a(1)$ level. As a consequence, after a few excitation-decay cycles, essentially all the molecules will end up into the a(1) level. A pump laser pulse at 1073 nm can now be used to excite the molecules from the a(1) level to the c(0) state. From the c(0), about 95\% of the molecules decay to the a(0) state and are tagged. A probe laser pulse at 905 nm can then be used to drive only the tagged $a(0)$ molecules to the $d$ state to produce the fluorescence. Compared to the old tagging method, this new scheme makes use of the $a(0)$ state to tag the molecules. The tagging efficiency is now 95\% instead of 4\%. Furthermore, for the tagged molecules in the a(0) state, multiple 905 nm laser pulses can be used to drive cycling transitions to enhance the fluorescence strength. Experimentally, one may pass the fluid seeded with the a(0) molecules through a region illuminated by the 910 nm pre-pumping laser so as to convert the $a(0)$ molecules into $a(1)$¡¯s (see Fig. \ref{Fig4} (b)). A focused pump laser pulse at 1073 nm tags a line of a(0) molecules which is then imaged by an expanded 905 nm probe laser pulse at a delayed time. Due to the enhanced tagging and imaging efficiency, it is expected that the overall signal-to-noise ratio can be improved by two orders of magnitude compared to that in the old tagging experiment.

\subsection{He$^*_2$ cloud tracking velocimetry}
In classical fluid dynamics research, quantitative PIV and PTV techniques are more widely used compared to MTV measurement \cite{Raffel-book-2007}. Unlike standard MTV where the velocity can be measured only at the locations of the tracer lines, PIV can generate a smoothly varying velocity field in the full visualization space, and PTV allows the measurement of Lagrangian quantities such as the local velocity and its derivatives. In He II, PIV and PTV measurement techniques have been developed based on the use of micron-sized tracer particles \cite{Sciver-JLTP-2007,Zhang-NP-2005,Mantia-JFM-2013,Paoletti-JPSJ-2008,Bewley-Nat-2006,Paoletti-PRL-2008}. However, the injection of these particles and their interactions with both fluid components in He II lead to known inherent issues as discussed in Sec. \ref{intro}. Therefore, it is natural to ask whether it is possible to perform PIV and PTV measurements in He II with smaller tracer particles, such as He$^*_2$ molecules. Unfortunately, the cycling transition laser-driven fluorescence scheme for detecting He$^*_2$ molecules has not yet been pushed to the limit for imaging individual He$^*_2$ molecules. Instead of tracking individual He$^*_2$ molecules, a more feasible route is to develop a method to produce many small clouds of He$^*_2$ molecules in He II and then treat each cloud as a ``tracer''. Such tracer clouds can be readily imaged for traditional PIV and PTV measurements.

\begin{figure}
\center
  \includegraphics[width=0.8\linewidth]{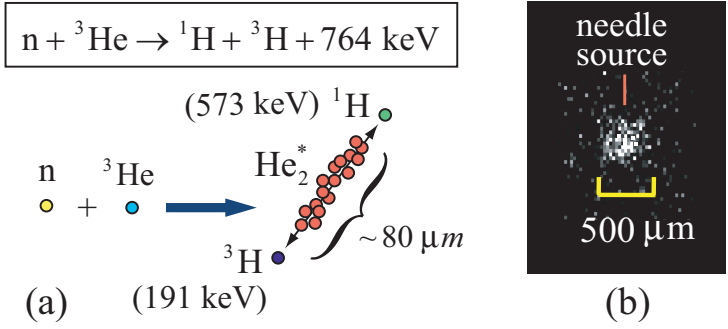}
\caption{(a) Schematic diagram showing the neutron-$^3$He absorption reaction that leads to the creation of a cloud of about $10^4$ He$^*_2$ molecules. (b) A fluorescence image of a cloud of about $10^3$-$10^4$ He$^*_2$ molecules, produced by applying a voltage pulse (30-60 ms) to a sharp tungsten needle in He II \cite{Guo-PRL-2009}.}
\label{Fig5}       
\end{figure}

To create small clouds of He$^*_2$ molecules in He II, an innovative method based on neutron-$^3$He absorption reactions was first proposed by the author at a workshop on quantum turbulence held in Abu Dhabi in 2012 \cite{Guo-talk-2012}. Then, this method was discussed in more details later at a neutron workshop at Oak Ridge National Lab in 2016 \cite{Guo-talk-2016}. The basic concept is that $^3$He atoms that naturally exist in He II as impurity particles can absorb neutrons as shown in the reaction schematic in Fig. \ref{Fig5} (a) \cite{Hayden-PRL-2014}. 
The absorption cross-section $\sigma$ depends on the energy of the incident neutrons and can be quite large for thermal neutrons that have spins aligned with the $^3$He atoms (i.e., $\sigma\sim10^{-20}$ cm$^2$) \cite{King-PR-1949,Meyer-JLTP-1997}. This reaction produces two charged particles, a proton ($^1$H) and a tritium ($^3$H), that move back to back with a total kinetic energy of 764 keV \cite{Meyer-JLTP-1997}. The two particles have sufficient energies to ionize and excite ground state $^4$He atoms along their tracks in He II, which leads to the generation of He$^*_2$ molecules. The total length of the two tracks, evaluated based on the known energy deposition rate of $^1$H and $^3$H in helium \cite{Meyer-JLTP-1997}, is about 80 $\mu$m. We can also estimate that about $10^4$ He$^*_2$ molecules are produced, considering the fraction of the kinetic energy of $^1$H and $^3$H that goes to the generation of He$^*_2$ molecules \cite{Guo-PRD-2013,Ito-PRC-2013}. Therefore, every $n$-$^3$He absorption event in He II creates a cloud of about 10$^4$ He$^*_2$ molecules with a size of about 80 $\mu$m. Imaging these clouds should be feasible since we have already demonstrated high-quality fluorescence imaging of tiny clouds of He$^*_2$ molecules created by pulsing a tungsten needle in He II (see Fig. \ref{Fig5} (b)) \cite{Guo-PRL-2009}. Due to the very small molecular diffusion of He$^*_2$ molecules in He II (i.e., $\sim10$ $\mu$m during typical drift time \cite{McKinsey-PRL-2005}), every cloud can be treated as a single tracer, and a spatial resolution of a few tens of microns is achievable. Tracking these He$^*_2$ clouds should allow us to map out the full-space velocity field.

\begin{figure}
\center
  \includegraphics[width=0.6\linewidth]{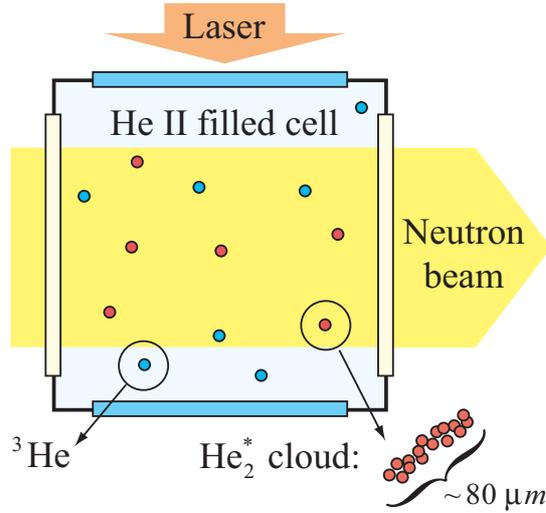}
\caption{Schematic of the experimental setup for producing and imaging He$^*_2$ clouds produced via $n$-$^3$He absorption reactions in He II.}
\label{Fig6}       
\end{figure}

Experimentally, one can pass a thermal neutron beam with a suitable flux (i.e., $\sim10^4$/cm$^2$ per pulse) through a He II filled sample cell (see Fig. \ref{Fig6}). By adjusting the concentration of the $^3$He impurity, one can conveniently tune the number density of the resulted He$^*_2$ clouds. For instance, with a $^3$He concentration as low as 50 ppm (part per million), which is far smaller than the number density of thermal rotons in He II above 1 K, we estimate that about $10^2$ clouds will be produced per 1 cm$^3$ volume of He II, which should be sufficient for velocity-field mapping purpose. This method indeed also leads to a very sensitive mechanism for neutron detection. Recently, two collaborative experimental projects have been launched on fluorescence detection of $n$-$^3$He absorption events in He II. The first project was initiated by Shimizu and colleagues at Nagoya University in collaboration with some researchers at other institutes in Japan and the author \cite{JPARC-collaboration}. This project utilizes the neutron facilities at the Japan Proton Accelerator Research Complex (J-PARC), and its first report has been submitted as a proceeding paper to the 2018 International Conference on Quantum Fluids and Solids (i.e., QFS2018) \cite{Matsushita-JLTP-2018}. The other project is being conducted at the Oak Ridge National Lab (ORNL) in the United State and is led by Fitzsimmons in collaboration with his colleagues at University of Tennessee and ORNL together with the author \cite{ORNL-collaboration}. Preliminary results obtained in both projects have shown promising evidences for He$^*_2$ cloud production in He II via $n$-$^3$He absorption reactions.

\section{Conclusion}\label{sec:3}
We have briefly reviewed the progresses made in recent years in developing quantitative flow visualization techniques applicable to He II using He$^*_2$ molecular tracers. A MTV technique based on tracking a thin line of He$^*_2$ molecules created via femtosecond laser-field ionization proves to be remarkably valuable. Nevertheless, two limitations of this method has been identified. These limitations are inherent of the single-line tracking scheme. We have discussed that by creating multiple tracer lines that form a tracer-line grid, one can perform stereoscopic MTV measurements that are completely immune from these limitations. Furthermore, two novel schemes for tagging He$^*_2$ tracers are discussed. The first method involves an inverse tagging scheme that greatly improves the tagging efficiency. It is expected that this method will allow the tagging and imaging of a He$^*_2$ tracer line without the use of any expensive femtosecond lasers. The second method relies on the creation of small clouds of He$^*_2$ tracers via $n$-$^3$He absorption reactions in He II. This method enables unambiguous quantitative PTV/PIV measurements of the full-space normal-fluid velocity field in He II.

The quantitative MTV techniques we have discussed are applicable not only to He II but also to the classical phase of liquid helium (He I) and gaseous helium. They have the potential to unlock the full power of cryogenic helium. For instance, helium gas is particularly useful in natural thermal convection research. Heat transfer and fluid flow in natural convection are controlled by the Rayleigh number $Ra$ and the Prandtl number $Pr$ \cite{Kakac-book-2013}. Studying the scaling laws of the flow parameters at large $Ra$ numbers, where turbulent convection sets in, is of both theoretical and practical significance \cite{Niemela-JLTP-2006}. Close to its critical point, gaseous helium can be used to produce convection flows with extremely large $Ra$, and the range of $Ra$ can be tuned by a factor of $10^{12}$ by simply adjusting the gas pressure \cite{Niemela-Nature-2000,Roche-NJP-2010,Urban-PRL-2012}, which makes laboratory study of ocean convection and atmospheric circulation possible \cite{Niemela-JLTP-2006}. However, velocity field measurement in helium gas has not been conducted so far, largely due to the lack of appropriate tools. The development of novel He$^*_2$ based quantitative MTV techniques may open the door for new avenues of research in this field.

\begin{acknowledgements}
 The author would like to acknowledge the contributions made by previous and current students in the lab, including J. Gao, A. Marakov, E. Varga, B. Mastracci, S. Bao, Y. Zhang, and H. Sanavandi. The author would also like to thank many colleagues in quantum turbulence and classical fluid dynamics research fields for valuable discussions. The work has been supported by U.S. Department of Energy under grant No. DE-FG02-96ER40952 and by the National Science Foundation under Grants No. DMR-1807291 and No. CBET-1801780. All the experiments have been conducted at the National High Magnetic Field Laboratory, which is supported by NSF Grant No. DMR-1644779 and the state of Florida.
\end{acknowledgements}

\end{document}